\documentstyle[12pt]{article}
\begin{document}
\def\a{\alpha}
\def\b{\beta}
\def\ch{\chi}
\def\d{\delta}
\def\e{\epsilon}
\def\E{{\cal E}}
\def\f{\phi}
\def\g{\gamma}
\def\h{\eta}
\def\et{\tilde{\eta}}
\def\i{\iota}
\def\j{\psi}
\def\k{\kappa}
\def\l{\lambda}
\def\m{\mu}
\def\n{\nu}
\def\o{\omega}
\def\p{\pi}
\def\q{\theta}
\def\r{\rho}
\def\s{\sigma}
\def\t{\tau}
\def\u{\upsilon}
\def\x{\xi}
\def\z{\zeta}
\def\D{\Delta}
\def\F{\Phi}
\def\G{\Gamma}
\def\J{\Psi}
\def\L{\Lambda}
\def\O{\Omega}
\def\P{\Pi}
\def\S{\Sigma}
\def\U{\Upsilon}
\def\X{\Xi}
\def\T{\Theta}
\def\vf{\varphi}
\def\ve{\varepsilon}
\def\cC{{\cal P}}
\def\cD{{\cal Q}}

\def\pp {\partial }
\def\pb {\bar{\partial }}
\def\ET {\tilde{E }}
\def\be{\begin{equation}}
\def\ee{\end{equation}}
\def\ben{\begin{eqnarray}}
\def\een{\end{eqnarray}}

\hsize=16truecm
\addtolength{\topmargin}{-0.6in}
\addtolength{\textheight}{1.5in}
\vsize=26truecm
\hoffset=-.6in
\baselineskip=7 mm

\thispagestyle{empty}
\begin{titlepage}

\begin{center}
 {\large\bf Conservation Laws in
 Higher-Order Nonlinear Optical Effects  }
\vglue .5in
Jongbae Kim\footnote{ Electronic address; jongbae@pathos.etri.re.kr }
\vglue .2in
  {\em Research Department, ETRI  \\  Yusong P.O. Box 106, Taejon 305-600,  Korea}
\vglue .2in
{and}
\vglue .2in
Q-Han Park\footnote{ Electronic address; qpark@nms.kyunghee.ac.kr },~~
H. J. Shin\footnote{ Electronic address; hjshin@nms.kyunghee.ac.kr }
\vglue .2in
{\it
Department of Physics  \\
Kyunghee University     \\
Seoul 130-701, Korea}
\vglue .2in
{\bf ABSTRACT}\\[.2in]
\end{center}
Conservation laws of the nonlinear Schr\"{o}dinger equation are studied in
the presence of higher-order nonlinear optical effects including
the third-order dispersion and the self-steepening.
In a context of group theory, we derive a general expression for
infinitely many conserved currents and charges of the
coupled higher-order nonlinear Schr\"{o}dinger equation.
The first few currents and charges are also presented  explicitly.
Due to the higher-order effects, conservation laws of the nonlinear Schr\"{o}dinger
equation are violated in general.
The differences between the types of the conserved currents for the Hirota and the
Sasa-Satsuma equations imply that the higher-order terms determine the inherent types of conserved quantities for
each integrable cases of the   higher-order nonlinear Schr\"{o}dinger equation.

\end{titlepage}

\newpage
In the ultrafast optical signal system, the higher-order nonlinear effects such
as the third-order dispersion, the self-steepening, and the self-frequency shift become
important if the pulses are shorter than $T_{0} \le 100 fs$ \cite{Agr}.
The use of optical pulses with distinct polarizations and/or frequencies also
require the consideration of nonlinear cross-couplings between different modes of pulses.
Inclusion of both the higher-order and the cross-coupling effects lead to the
study on the coupled higher-order nonlinear Schr\"{o}dinger equations (CHONSE)
which are not in general integrable except for special cases of coupling constants.
Those integrable cases of coupling constants have been classified
in association with  Hermitian symmetric spaces \cite{Park}.
It is also well known that soliton equations which can be integrated by the inverse scattering
method possess infinite number of conserved quantities. For example, the nonlinear
Schr\"{o}dinger equation (NSE)  has infinite number of conserved charges in addition to the  ones
corresponding to the energy and the intensity-weighted mean frequency. However, the effect
of the higher-order and the cross coupling terms on the conservation laws has not been considered
up to now.

In this paper, we make a systematic study on  the conservation laws in the presence of
the higher-order and the cross-coupling terms.
We first indicate that except for
the energy conservation, other conservation laws of the NSE such as the conservation of
the intensity-weighted mean frequency do not hold due to the higher-order effects any more,
unless the higher-order terms are of a unique type.
In the  case of integrable CHONSE, we derive general expressions of
 infinite number of conserved currents and charges from
the Lax pair formulation utilizing the properties of the Hermitian symmetric space.
From the general expressions,  explicit forms of the first few conserved
currents and the associated charges of the Hirota and the Sasa-Satsuma equations
are  calculated.
We then explain the correlations of conservation laws between the two integrable cases
of the higher-order extension of the NSE.

In order to illustrate the issue, we first consider the NSE including the higher-order
terms. In a mono-mode optical fiber, the propagation of a ultrashort pulse is governed by
the higher-order nonlinear Schr\"{o}dinger equation \cite{hnls1985}
\be
\pb \j = i (\gamma_1 \pp^2 \j + \gamma_2 |\j|^2 \j ) +
           \gamma_{3}\pp^3 \j + \gamma_4 \pp (|\j|^2 \j) + \gamma_5 \pp (|\j |^2 ) \j,
\label{hnls}
\ee
where $\pb \equiv \pp / \pp \bar{z}$ and $\pp \equiv \pp / \pp z$ are derivatives
in retarded time coordinates ($\bar{z} =x, z=t-x/v $), and $\j$
is the slowly varying envelope function.
The real coefficients $\gamma_{i}$ $(i=1,2,3,4)$ in the first four terms
on the right hand side of Eq. (\ref{hnls}) specify in sequence the effects of the
group velocity dispersion, the self-phase modulation, the third order dispersion,
and the self-steepening. With appropriate scalings of space, time, and field variables,
one can readily normalize Eq. (\ref{hnls}) so that $\gamma_1 = 1 , ~ \gamma_2
=2,~ \gamma_3=1$ which we assume from now on.
The remaining coefficient $\gamma_{5}$ in the last term is complex in general.
The real and the imaginary parts of $\gamma_{5}$ are due to the effect of the
frequency-dependent radius of fiber mode and the effect of the self-frequency shift
by stimulated Raman scattering, respectively.
It is well known that the above equation becomes integrable if $\gamma_4 = -\gamma_5 =6$
(Hirota case) \cite{Hirota} or $\gamma_4 = -2\gamma_5 =6$ (Sasa-Satsuma case) \cite{Sasa}.
In the absense of higher-order terms ($\gamma_3 = \gamma_4 = \gamma_5 =0$),
Eq. (\ref{hnls}) possesses infinite number of conserved charges among which the first
three charges  \cite{Hase1989} are
\ben
Q_{1} &=&  \int_{-\infty}^{ \infty} |\j |^2 dt  ~     ,
\nonumber \\
Q_{2} &=&  i \int_{-\infty}^{ \infty} (\j^{*} \pp \j - \pp \j^{*} \j ) dt  ~ ,
\nonumber \\
Q_{3} &=&  \int_{-\infty}^{ \infty} (\pp \j^{*} \pp \j - |\j |^4 ) dt,
\label{3charges}
\een
where $Q_{1} $ represents conserved energy, and $Q_{2}$ the mean frequency weighted by
the intensity of optical pulses. In the conventional  NSE
where the time and the space coordinates are interchanged, $Q_{1}, Q_{2}$ and $Q_{3}$
respectively correspond to conserved mass, momentum and energy. If we include
higher-order terms, $Q_{i}$ are not necessarily conserved but subject to
the relations;
\ben
\pb Q_{1} &=& 0     ~ ,
\nonumber \\
\pb Q_{2} &=& 2 i (\g_{4} +\g_{5})\int_{-\infty}^{ \infty}\pp |\j |^2       ~ ,
(\j^{*} \pp \j - \pp \j^{*} \j ) dt
\nonumber \\
\pb Q_{3} &=&  (3 \g_{4} +2 \g_{5} -6 )\int_{-\infty}^{ \infty}\pp |\j |^2
\pp \j^{*} \pp \j dt      ~ .
\label{consnls}
\een
The calculations indicate that the charge $Q_{1}$
which corresponds to energy is conserved for all
values of $\g_{4}$, $\g_{5}$ while  $Q_{2}$ and $Q_{3}$
are conserved provided $\g_{4} + \g_{5} =0$ and
 $ 3 \g_{4} +2 \g_{5} = 6$, respectively. Note that $Q_{2}$ and $Q_{3}$  are
 conserved simultaneously only for the specific value $\g_{4} = -\g_{5} =6$ that  is
precisely the Hirota case.
It is interesting to observe that integrability does
not always imply  the same types of conserved currents
in the presence of higher-order terms.
 Another integrable case  of  the Sasa-Satsuma equation,
where $ \gamma_4 = -2\gamma_5 =6$,  in fact does not have
$Q_{2}$ and $Q_{3}$ in Eq. (\ref{consnls})
as the conserved charges. This consequence is rather remarkable in view of the fact that integrable
equations possess infinite number of conserved quantities. We will show, however,
the Sasa-Satsuma equation also possesses infinitely  many conserved charges
of different types other than the ones of the Hirota equation.

 In case we include both the higher-order and the cross-coupling nonlinear effects,
the propagating system is governed by a CHONSE.
Without understanding  physical settings, it would be meaningless to
write down any general expression of
 the CHONSE. However, as explicitly derived in \cite{Park}, there
exists a  group theoretic specification  which admits a systematic  classification
of integrable cases of the CHONSE. In the following, we consider a group theoretic
generalization of the NSE and define the CHONSE in association with a
Hermitian symmetric space.
By solving the linear Lax equations  iteratively,
we derive infinite number of conserved currents and charges for the CHONSE.
For the later use, now we briefly review the definition of  Hermitian symmetric spaces
\cite{Fordy,Hel}
and the generalization of the NSE  \cite{Park,Oh} according to
the Hermitian symmetric spaces.

A symmetric space is a coset space $G/K$ for Lie groups $G \supset K$
whose associated Lie algebras ${\bf g}$ and $ {\bf k}$, with the
decomposition:
${\bf g} = {\bf k} \oplus {\bf m}$, satisfy the commutation
relations;
\begin{equation}
[{\bf k} , ~ {\bf k}] \subset {\bf k}, ~~ [{\bf m},~ {\bf m}] \subset {\bf k},  ~~
[{\bf k}, ~ {\bf m}] \subset {\bf m}
\label{algebra}
\end{equation}
A Hermitian symmetric space is the symmetric space $G/K$ equipped with a complex
structure. One can always find an element $T$ in the Cartan subalgebra of
${\bf g}$ whose adjoint action defines a complex structure and also the
subalgebra ${\bf k}$ as a kernel, i.e.,  ${\bf k} = \{ V \in {\bf g}:~ [T,~ V] = 0 \}$.
That is, the adjoint action  $J \equiv \mbox{ad}T = [T, ~ *]$ is a linear map
$J: {\bf m} \rightarrow {\bf m}$ that satisfies the complex structure condition,
$J^{2} = -I $,  or $[T, ~ [T,~ M]] = - M $ for $ M \in {\bf m}$.
Then, we define a CHONSE as\footnote{We restrict to symmetric spaces $AIII, CI$ and $DIII$
only so that the expression of CHONSE becomes simplified \cite{Park}. }
\be
\pb E = \pp^2 \ET -  2 E^2 \ET + \a ( \pp^3 E +\b_{1} E^2 \pp E +\b_{2} \pp E E^2 )
\label{chonse}
\ee
where $E$ and $\ET \equiv [T, ~ E]$ are  extended field variables  belonging to ${\bf m}$.
The arbitrary constant $\a $ may be normalized to 1 by an appropriate
scaling but we keep it in order to exemplify the higher-order effects.
Also the cross-coupling effects between different modes of polarizations or frequencies
are accommodated in the matrix form of $E$ which is determined by  each Hermitian
symmetric space.
For example, in the case where  $G/K=SU(N+1)/U(N)$, matrices $E$ and $T$ are represented as
\be
E = \pmatrix{0 & \j_1 & \cdots & \j_N \cr -\j_1^* & 0 & & 0 \cr \vdots & && \vdots \cr
             -\j_N^* & 0 & \cdots & 0}, ~~
T=\pmatrix{{i\over 2} & 0 & \cdots & 0 \cr 0 & -{i\over 2} & & 0\cr &  & \ddots & \cr
0 & \cdots & 0 &-{i \over 2}}  ,
\label{cpn}
\ee
and the CHONSE becomes the higher-order vector nonlinear Schr\"{o}dinger equation,
\ben
\pb \j_{k} &=& i \Big[ \pp^2 \j_{k} + 2(\sum_{j=1}^{N}|\j_{j}|^2) \j_{k} \Big]
\nonumber \\
           &-& \alpha \Big[ \beta_{1}(\sum_{j=1}^{N}|\j_{j}|^2) \pp \j_{k} +
            \beta_{2} (\sum_{j=1}^{N}\j^{*}_{j} \pp \j_{j}) \j_{k} - \pp^3 \j_{k} \Big]
            ~; ~ k=1,2, ..., N .
\label{hvnls}
\een
This equation is an obvious generalization of Eq. (\ref{hnls}) to the multi-component case.
It is easy to see that Eq. (\ref{hvnls}) with $N=1$ and $\b_{1} = \b_{2} = -3$
is precisely the Hirota equation.
As verified in \cite{Park}, Eq. (\ref{chonse}) is integrable if $\b_{1} = \b_{2} = -3$
because in such a case the CHONSE admits a Lax pair.
That is, Eq. (\ref{chonse}) with $\b_{1} = \b_{2} = -3$ arises
from the compatibility condition ( $[L_{z} , ~ L_{\bar{z}}]=0$) of the
associated linear equations,
\ben
L_{z} \Psi  & \equiv & [ \partial + E +\lambda T ] \Psi  = 0 , \nonumber \\
L_{\bar{z} } \Psi & \equiv & [ \pb +U^{0}_{K}+U^{0}_{M}+\lambda(U^{1}_{K}+U^{1}_{M})
                 +\lambda^{2}(U^{2}_{M}+ T)-\alpha \lambda^{3} T] \Psi
= 0 ,
\label{Lax}
\een
which holds for all values of the spectral parameter $\l $.
The entities $U^{i}_{K}$ and $U^{i}_{M}$ in $L_{\bar{z} }$ are
given by
\ben
U^{0}_{K} &=&  -E \ET  -\a [ E , \pp E ] , ~~
U^{0}_{M} = \pp \ET + \a (\pp^{2} E -2 E^3) , \nonumber \\
U^{1}_{K} &=& {\a} E\ET , ~~ U^{1}_{M}  =   E -\a \pp \ET ,
~~ U^{2}_{M}   = - \a E .
\een
Here, the subscripts $K$ and $M$ signify that they belong to the subalgebra ${\bf k}$
and the remaining complement ${\bf m}$,  respectively.
The algebraic decomposition can be also extended to a more general case
including the matrix  solution $\Psi = \Psi_{K} + \Psi_{M} $
 with the properties that $[T, \Psi_{K}] =0, ~ [T, \Psi_{M}]  \in {\bf m}$,
 and  the following multiplication properties;
\be
[T, ~ \Psi^{1}_{K}\Psi^{2}_{K} ] = [T, ~  \Psi^{1}_{M}\Psi^{2}_{M} ] =0, ~
[T, ~ \Psi^{1}_{K}\Psi^{2}_{M}]  \in {\bf m}
\label{decom}
\ee
The adjoint action of the element $T$ in the Cartan subalgebra together with the complex
structure condition, if applied to the  decomposition,
lead to  a couple of general identities for any $ M_{1}, M_{2} \in {\bf m}$;
\be
   [T, ~ M_{1}M_{2}]  = \tilde{M}_{1} M_{2} + M_{1} \tilde{M}_{2} = 0 , ~~~
   \tilde{M}_{1} \tilde{M}_{2} = M_{1} M_{2}
\label{ident}
\ee
These identities are useful for  many calculations, for example,  in deriving conserved currents
or   in verifying that the CHONSE in Eq. (\ref{chonse}) is equivalent
to the compatibility condition of the Lax pair in Eq. (\ref{Lax}).

Having presented necessary ingredients, we are now ready to derive infinitely many
conserved currents and charges of the integrable CHONSE
by solving the associated linear equations
in Eq. (\ref{Lax}). In order to make use of the algebraic properties of Hermitian
symmetric spaces, we make a change of  the variable $\Psi$ in Eq. (\ref{Lax})
by
\be
\Phi =  \Psi \exp[( \lambda z + (\lambda^{2} - \a \lambda^{3})\bar{z})T] ,
\label{PsiPhi}
\ee
which results in the change of the multiplicative term $T\Psi $ to the commutative term
$  [T, \Phi]$ in the linear equations.
 The adjoint action, $ [T, \Phi]$, allows the splitting of the linear equations for $\Phi$
into the $K$- and the $M$-components as explained below.
 Let us first assume that the linear equations
can be solved iteratively in terms of
\be
\Phi (z, \bar{z} , \l ) \equiv  \sum_{n=0}^{\infty}\frac{1}{\lambda^{n}}\Big(
\Phi_{K}^{n}(z, \bar{z})
+\Phi_{M}^{n}(z, \bar{z} ) \Big) ,
\label{solPhi}
\ee
where $\Phi_{K}^{n}$ and $\Phi_{M}^{n}$ denote  the  decomposition  of  a coefficient
$\Phi^{n}$ satisfying the properties in Eq. (\ref{decom}).
Then, the n-th order equation ($n \ge 0$) separates into the $K$- and the $M$-components
such as
\ben
  & &   \partial\Phi_{K}^{n} + E \Phi_{M}^{n} =0 ,
\label{dK}            \\
   & &  \partial\Phi_{M}^{n} +  E \Phi_{K}^{n} + [T, \Phi_{M}^{n+1}]  =0 ,
\label{dM}
\een
 while the $\pb$-part of the linear equation becomes
\ben
   & & \overline{\partial}\Phi_{K}^{n} + U^{0}_{K}\Phi_{K}^{n}+U^{0}_{M}\Phi_{M}^{n}
          +U^{1}_{K}\Phi_{K}^{n+1}+U^{1}_{M}\Phi_{M}^{n+1}+U^{2}_{M}\Phi_{M}^{n+2} =0 ,
\label{dbarK}            \\
 & &  \overline{\partial}\Phi_{M}^{n} + U^{0}_{K}\Phi_{M}^{n}+U^{0}_{M}\Phi_{K}^{n}
   +U^{1}_{K}\Phi_{M}^{n+1}+U^{1}_{M}\Phi_{K}^{n+1}+U^{2}_{M}\Phi_{K}^{n+2}
    +[T, \Phi_{M}^{n+2}]- \alpha [T, \Phi_{M}^{n+3}] =0 .
    \nonumber    \\
\label{dbarM}
\een
In addition, there are equations arising from the positive powers
of $\l$, which can be  given by Eqs. (\ref{dK})-(\ref{dbarM}) provided  that
$n=-1, -2, -3$ and $\Phi_{K}^{n < 0}=\Phi_{M}^{n < 0}=0$ are defined.
These equations can be solved recursively for  $\Phi_{K}^{n}$, $\Phi_{M}^{n}$ $(n \geq 0)$
starting from a consistent set of initial conditions;
\be
\Phi_{M}^{0}  =  0, ~~ \Phi_{K}^{0}  =  -i I    ,~~
     \Phi_{M}^{1}  =  -i \ET  .
  \label{inicon}
\ee
Note that Eq. (\ref{dM}) can be solved for $ \Phi_{M}^{n+1}$ by using the complex structure
condition.  That is, $\Phi_{M}^{n+1}= - [T, ~[T, ~ \Phi_{M}^{n+1}]]=
[T, \partial\Phi_{M}^{n}]+ \ET \Phi_{K}^{n}$.
Thus, $\Phi_{M}^{n+1}$ is obtained directly provided that $\Phi_{K}^{n}$ and $\Phi_{M}^{n}$
are determined. Contrary to $\Phi_{M}^{n+1}$ which is obtained algebraically,
$\Phi_{K}^{n+1}$ can be obtained by a direct integration of Eq. (\ref{dK}). In fact,
$\Phi_{K}^{n+1}$ is over-determined due to Eq. (\ref{dbarK}) as well. Thus, in order for
$\Phi_{K}^{n+1}$ to be integrable, the compatibility condition that
$[ \pp,  \pb] \Phi^{n}_{K} = 0$ should be required. The condition is
satisfied provided  the integrable CHONSE holds. In this case
the compatibility condition   gives rise to
infinitely many conserved currents labeled by integer $n$ such that
$\pp \bar{J}_{K}^{n} +  \pb J_{K}^{n} =0$;
\ben
J_{K}^{n} &=&  - \pp \Phi_{K}^{n} =  E \Phi_{M}^{n}
            \\
\bar{J}_{K}^{n} &=& \pb \Phi_{K}^{n} =
          -( \pp \ET + \a \pp^{2} E  - 3\a E^{3}) \Phi_{M}^{n}- \a E \pp^2 \Phi_{M}^{n}
  - (E - \a \pp \ET ) [T, \pp \Phi_{M}^{n}]
\een
In order to derive the local currents  explicitly, we solve the recurrence relations in
Eqs. (\ref{dK})-(\ref{dbarM}) with the initial conditions as in
Eq. (\ref{inicon}). The first few conserved currents are listed below;
\ben
J_{K}^{1} &=&  -i  E \ET                             ,
\nonumber \\
\bar{J}_{K}^{1} &=&  -  i[E, \pp E] +
i \a ( [\pp^2 E, \ET] + \pp  \ET \pp E  - 3 E^3 \ET ) ,
\een
for $n=1$, and
\begin{eqnarray}
J_{K}^{2} &=& - i \pp \Phi^{1}_{K} \Phi^{1}_{K} + i E \pp E  ,
\nonumber \\
\bar{J}_{K}^{2} &=& +i \pb \Phi^{1}_{K} \Phi^{1}_{K}
 - i (E \pp^2 \ET +\pp \ET \pp E  - E^{3}\ET )
\nonumber \\
&& - i \a ( E\pp^3 E + [\pp^2 E, \pp E] + \pp E E^{3} -2 E \pp E E^2 -E^2 \pp E E -4 E^3 \pp E) ,
 \\
J_{K}^{3}
&=& - i(\pp \Phi^2_{K} \Phi^1_{K} +  \pp \Phi^1_{K} \Phi^2_{K}
-i \pp \Phi^1_{K} \Phi^1_{K} \Phi^1_{K} ) +i (E \pp^2 \ET - E^3 \ET) ,
\nonumber \\
\bar{J}_{K}^{3} &=&  i( \pb \Phi^1_{K} \Phi^2_{K} + \pb \Phi^2_{K} \Phi^1_{K}
                        -i  \pb \Phi^1_{K} \Phi^1_{K} \Phi^1_{K} )
\nonumber \\
& + &  i(E \pp^3 E -\pp E \pp^2 E +\pp E E^3 - 2 E \pp E E^2 - E^2 \pp E E - 2E^3 \pp E)
\nonumber \\
& - & i \a (E \pp^4 \ET + \pp^2 E \pp^2 \ET +  \pp\ET \pp^3 E- 5 E^3 \pp^2 \ET -E^2 \pp^2 E \ET
-3E \pp^2 \ET E^{2}
   \nonumber \\
& &   +\pp^2 \ET E^3   -2 \pp \ET \pp E E^2
 -\pp \ET E \pp E E - 2\pp \ET E^2 \pp E
  -3 E \pp E \pp E \ET
\nonumber \\
& & -5 E \pp E E \pp \ET  -3 E^2 \pp E \pp \ET
+ 4 E^5 \ET ) ,
\end{eqnarray}
for $n=2$ and $n=3$, respectively.
Note that currents $J^{n}_{K} $ and $\bar{J}^{n}_{K} $ for $n \geq 2 $
contain non-local terms $\Phi^{m}_{K}$ with  $ m < n$.
Fortunately, these non-local terms can be separated from the conservation law
if we consider a scalar expression of the conserved current by
taking an appropriate trace as follows;
\be
S^{n}_{K} = \mbox{Tr} (P J^{n}_{K} ) , ~~
\bar{S}^{n}_{K} = \mbox{Tr} (P \bar{J}^{n}_{K} )
\ee
The parameter $P$ is any matrix entity  which commutes with matrices $\Phi^{m}_{K}$, or
we may choose  $P = c_{1} I + c_{2} T$ for arbitrary constants $c_{1} $ and $ c_{2}$.
For instance, we have for $n=2, 3$
\begin{eqnarray}
S_{K}^{2} &=& - \pp ( \mbox{Tr } P \{ {i \over 2}  (\Phi^{1}_{K})^2 \} )
+  \mbox{Tr } P ( i  E \pp E)   ,
\nonumber \\
\bar{S}_{K}^{2} &=& \pb ( \mbox{Tr }P \{  {i \over 2}  (\Phi^{1}_{K})^2 \} )
\nonumber \\
   & + & \mbox{Tr } P  \{ - i  (E \pp^2 \ET +\pp \ET \pp E  - E^{3}\ET )
- i \a ( E\pp^3 E + [\pp^2 E, \pp E] -6 E^3 \pp E)
\}                               ,
    \\
S_{K}^{3}
&=& - \pp ( \mbox{Tr } P \{  i ( \Phi^1_{K} \Phi^2_{K}
-{1\over 3} ( \Phi^1_{K})^3 )\} )  +\mbox{Tr } P \{ i (E \pp^2 \ET - E^3 \ET) \} ,
\nonumber \\
\bar{S}_{K}^{3} &=&  \pb ( \mbox{Tr }  P \{  i ( \Phi^1_{K} \Phi^2_{K}
-{1\over 3} ( \Phi^1_{K})^3 )\} )  ,
\nonumber \\
 &+&  \mbox{Tr } P \{ i (E \pp^3 E -\pp E \pp^2 E   - 4E^3 \pp E)
   - i \a (E \pp^4 \ET + \pp^2 E \pp^2 \ET +  \pp\ET \pp^3 E
- 8 E^3 \pp^2 \ET
\nonumber \\
 & &
+ 2\pp^2 \ET E^3  + E^2\pp \ET\pp E
 -\pp \ET E \pp E E +\pp \ET E^2 \pp E -5 E \pp \ET  E  \pp E+ 4 E^5 \ET )     \}
 .
\end{eqnarray}
The derivations show that the non-local terms appear as total derivative terms thus they are
conserved separately. Dropping the non-local terms and integrating
over the time coordinate,  we obtain infinite number of
global charges which are conserved in space, i.e. $\pb Q^{n}_{K} =0$
where
\be
Q_{K}^{n} \equiv \int_{-\infty}^{+\infty}dt S_{K}^{n}   .
\label{localch}
\ee
For the case of $G/K=SU(N+1)/U(N)$ as mentioned in Eq. (\ref{cpn}), we work out explicitly
and obtain the conserved charges
 \be
    Q_{K}^{1} =  \int_{-\infty}^{+\infty}dt  \sum_{k=1}^{N} \psi_{k}^{*}\psi_{k}   ,
    \label{HirotaQ1}
\ee
for $ n=1$ and
\begin{eqnarray}
  & & Q_{K}^{2}  = i \int_{-\infty}^{+\infty}dt \sum_{k=1}^{N}
              (\psi_{k}^{*} \pp \psi_{k} - \pp \psi_{k}^{*}\psi_{k}  )   ,
    \label{HirotaQ2}
     \\
  & & Q_{K}^{3} = \int_{-\infty}^{+\infty}dt [
              \sum_{k=1}^{N} \pp\psi_{k}^{*} \pp \psi_{k}
                      - ( \sum_{k=1}^{N}\psi_{k}^{*}\psi_{k})^{2}       ]     ,
    \label{HirotaQ3}
\end{eqnarray}
for $n=2$ and $n=3$, respectively.
Conserved charges for other cases of integrable CHONSE can be similarly obtained
from the specification of $E$ and $T$ as classified in \cite{Park}.

As  noted in Eq. (\ref{consnls}), the types of charges $Q_{2}$ and $Q_{3}$ are not
conserved in the Sasa-Satsuma case where $\gamma_4 = -2\gamma_5
=6$. Nevertheless, the Sasa-Satsuma equation equivalently possesses  infinitely
many  conserved charges of different types as well. This seemingly contradicting
characteristics can be explained by the fact that the Sasa-Satsuma equation arises
from the discrete $Z_{2}$-reduction of the $SU(3)/U(2)$ CHONSE  combined with a point
transformation \cite{Park}.
In this case, matrices $E$ and $T$ can be denoted as
\be
E = \pmatrix{0 & \j &  \j^{*} \cr -\j^* & 0  & 0 \cr
             -\j & 0 &  0}, ~~
T=\pmatrix{{i\over 2} & 0 & 0 \cr 0 & -{i\over 2}  & 0\cr
0 &  0 &-{i \over 2}} .
\label{Sasa}
\ee
Since the charge $Q^{n}_{K}$ in Eq. (\ref{localch}) is invariant under the point
transformation, we can also calculate  the first few conserved  charges of the
Sasa-Satsuma equation using the expressions $E$ and $\ET \equiv [T,~ E]$ given in
Eq. (\ref{Sasa}).  The resulting  charges of the Sasa-Satsuma equation are
\ben
    Q_{K}^{1} &=&   \int_{-\infty}^{+\infty}dt   \psi^{*}\psi     ,
\nonumber \\
  Q_{K}^{2}  &=& 0   ,
   \nonumber  \\
  Q_{K}^{3} &=& \int_{-\infty}^{+\infty} dt [3\partial\psi^{*}\partial\psi-6(\psi^{*}\psi)^{2}
                     - i (\psi^{*}\partial\psi-\partial\psi^{*}\psi) ]
.
  \label{3chargesSasa}
\een
  If the charges  in Eq. (\ref{3chargesSasa}) are compared with those of the Hirota type
  in  Eq. (\ref{3charges}) (or equivalently Eq. (\ref{HirotaQ1})-(\ref{HirotaQ3})
for $N = 1$),
we note that the current for $n=1$, which corresponds to energy,  is the same but
other currents are of different types.   Remarkably,
 in Eq. (\ref{3chargesSasa})  the current for $n=2$ turns out to be
trivial while the current for $n=3$ is a new type that is seemingly
combination of currents for $n=2, 3$
 in  Eq. (\ref{3charges}).
From Eq. (\ref{hnls}) with normalized coefficients $\g_{1}=\g_{2}/2=\g_{3}=1$, one
can readily confirm that the current
$S_{K}^{3}= 3\partial\psi^{*}\partial\psi-6(\psi^{*}\psi)^{2}
                     - i (\psi^{*}\partial\psi-\partial\psi^{*}\psi) $ is conserved only if
$\gamma_{4}+\gamma_{5}=3$ and $3\gamma_{4}+2\gamma_{5}=12 $.  Solving the
equations results in  $\g_{4}=-2\g_{5}=6$ that definitely leads to the Sasa-Satsuma case,  to be
compared with Eq. (\ref{consnls}) for the Hirota case.

To summarize, using the properties of Hermitian symmetric space
we have constructed the Lax pair formalism of the coupled higher-order nonlinear
Schr\"{o}dinger equation and derived a general expression of infinite number of
conservation laws.
Remarkably, the conserved currents and charges for both the Hirota and the Sasa-Satsuma
equations  are calculated from the general expressions,  accompanying the reduction procedure.
We have shown that, except for the Hirota case, the current conservations of the nonlinear
Schr\"{o}dinger equation are in general
broken by the higher-order effects.
The types  of conserved currents and charges for the Sasa-Satsuma case are different
from  the types for the Hirota case except for the energy conserved irrespective of
all the higher-order effects.
These differences may leave scope for  more physical explanations
and applications in  the further study of higher-order effects  including numerical analysis.
\vglue .2in
{\bf ACKNOWLEDGMENT}
\vglue .2in
J. Kim is supported by the Ministry of Information and Communication of Korea.
Q.H. Park and H.J. Shin are  supported in part by the program of Basic Science Research,
Ministry of Education BSRI-97-2442, and by Korea Science and Engineering
Foundation through CTP/SNU and  97-07-02-02-01-3.
\vglue .2in

\end{document}